\documentclass{ws-ijmpe}

\newcommand{\degree}{\ensuremath{^\circ}}

\begin{document}

\markboth{Rafael Alves Batista, Ernesto Kemp, Bruno Daniel}
{Detection of Point Sources in Cosmic Ray Maps Using the Mexican Hat Wavelet Family}

\catchline{}{}{}{}{}
%
%%%%%%%%%%%%%%%%%%%%%%%%%%%%%%%%%%%%%%%%%%%%%%%%%%%%%%%%%%%%%%%%%%%%

\title{DETECTION OF POINT SOURCES IN COSMIC RAY MAPS USING THE MEXICAN HAT WAVELET FAMILY}

\author{RAFAEL ALVES BATISTA, ERNESTO KEMP, BRUNO DANIEL}

\address{Instituto de Física ``Gleb Wataghin'', Universidade Estadual de Campinas\\ 
Cidade Universitária ``Zeferino Vaz'', 13083-859\\
Campinas-SP, Brazil\\
rab@ifi.unicamp.br}

\maketitle

\begin{history}
\received{Day Month Year}
\revised{Day Month Year}
\comby{Managing Editor}
\end{history}

\begin{abstract}
An analysis of the sensitivity of gaussian and mexican hat wavelet family filters to the detection of point sources of ultra-high energy cosmic rays was performed. A source embedded in a background was simulated and the number of events and amplitude of this source was varied aiming to check the sensitivity of the method to detect faint sources with low statistic of events.
\end{abstract}


\section{Introduction}

\par The origin of the Ultra-High Energy Cosmic Rays (UHECRs) is still an unsolved problem in astroparticle physics. A model for acceleration of cosmic rays, firstly devised by Hillas\cite{hillas}  predicts that cosmic rays are accelerated to the highest energies (above 10$^{18}$ eV) by electromagnetic fields of astrophysical objects.  Therefore, the identiﬁcation of possible astrophysical sources of UHECRs is possible by analysing the arrival directions of the cosmic rays. The correlation of the positions of point-like astrophysical objects (point sources) with the arrival directions of cosmic rays defines a small scale anisotropy. In the search of point-like sources it is a common procedure to convolve the sky maps containing arrival directions of cosmic rays with mathematical functions (the kernel of the convolution operation) aiming to optimize the signal to noise ratio. In this work it is studied the performance of some kernels of the Mexican Hat Wavelet Family (MHWF) to identify point sources of cosmic rays, and disentangle genuine signals from the background.

\section{Wavelets}

\par Wavelets are defined as mathematical functions belonging to the $\mathbb{L}^2$ space. They can be thought as localized wave-like oscillating functions which can be operated with a given signal and provide information about it. The continuous wavelet transform (CWT) in two dimensions may be formally written as
\begin{equation}
 \Phi(s,\tau_1,\tau_2) = \int \int f(t,u)\Psi^{*}_{s,\tau_1,\tau_2}(t)dtdu,
\end{equation}
where $s$ ($s>0$, $s {\ }\in {\ } \mathbb{R}$) is the scaling factor and $\tau_1$ and $\tau_2$ ($\tau_i {\ } \in {\ }\mathbb{R}$) are the translation parameters. So, the CWT decomposes a function $f(t,u)$ in a basis of wavelet $\Psi_{s,\tau_1,\tau_2}(t,u)$. One can scale and translate a ``mother-wavelet" $\Psi$ and obtain a wavelet $\Psi_{s,\tau_1,\tau_2}(t,u)$, as follows:
\begin{equation}
	\Psi_{s,\tau_1,\tau_2}(t,u)=\frac{1}{\sqrt{s}} \Psi\left( \frac{t-\tau_1}{s}, \frac{u-\tau_2}{s} \right).
\end{equation}

\par The Mexican Hat Wavelet Family (MHWF), introduced by Gonz{\'a}lez-Nuevo {\it et al}.\cite{mexicanhat}, and its extension on the sphere have been widely used to detect point sources in maps of cosmic microwave background radiation\cite{cayon,vielva01,vielva03}, due to the amplification of the signal-to-noise ratio (SNR) when transiting from real to wavelet space. The MHWF is obtained by successive application of the laplacian operator to the two-dimensional gaussian $\phi(\vec{x})$. A generic member of this family, of order $n$, is:
\begin{equation}
	\Psi_n(\vec{x}) = \frac{(-1)^n}{2^n n!} \nabla^{2n} \phi (\vec{x}).
\end{equation}

\section{Celestial Maps}

\par Celestial maps are pixelations of the celestial sphere taking into account the angular resolution of the experiment. The events map is a celestial map representing the arrival directions of cosmic rays in a suitable coordinate system. Due to intrinsic limitations of detector, every event detected is convolved with a probability related to the angular resolution of the detector, which means that there is a point spreading function (PSF) associated to the detector.

\par The convolution of celestial maps with filters is given by:
  \begin{equation}
  	M_f(k) = \frac{\sum_j M(j) \Phi(\vec{r_k},\vec{r_j})}{\sum_j \Phi(\vec{r_k},\vec{r_j})},
  \end{equation}
where $M(j)$ is the number of cosmic rays within the pixel of index $j$, in the direction $\vec{r_j}$. $\Phi(\vec{r},\vec{r_0})$ is the used filter and $\vec{r_k}$ is the position vector representing the point where the integral is being calculated.

\section{Analysis Procedure}

\par This work is an extension of previous ones\cite{physicae,icrc}. The simulated detector has two sites, one located in the southern hemisphere (36$\degree$ S and 65$\degree$ W), and the other in the northern hemisphere (38$\degree$ N and 102$\degree$ W), seven times larger than the one in the south, implying on a flux of cosmic rays seven times greater. The acceptance law has the form $sin\theta cos\theta$, where $0\degree\leq \theta \leq 60\degree$ is the zenith angle. It was also considered the case of an ideal detector with uniform exposure and full sky coverage.  

\par It was simulated a point  source located at (l,b)=(320$\degree$,30$\degree$) (galactic coordinates). This source was modeled by a gaussian:
\begin{equation}
	\label{eq:source}
	\frac{A}{2\pi}\exp\left( -\frac{\vec{x}^2}{2\sigma^2}  \right),
\end{equation}
and was embedded in a background consisting on a superposition of four different patterns of arrival directions of cosmic rays. The simulated patterns were: (i) an isotropic distribution of events; (ii),(iii) dipoles, modeled according to $\Phi(\hat{u})=\frac{\Phi_0}{4\pi} \left( 1+\alpha \hat{D}.\hat{u}  \right)$,
%\begin{equation}
%	\Phi(\hat{u})=\frac{\Phi_0}{4\pi} \left( 1+\alpha \hat{D}.\hat{u}  \right),
%\end{equation}
where $\Phi$ is the flux of cosmic rays, $\Phi_0$ is related to the isotropic flux, $0\leq \alpha \leq 1$ is the amplitude of the dipole ($\alpha=0.07$ for (ii) and $\alpha=0.005$ for (iii)), $\hat{D}$ is the vector which points towards the dipole ($(l,b)$ = $(0\degree,0\degree)$) for (ii)   and $(l,b)$ = $(166.5\degree,-29\degree)$) for (iii)) and $\hat{u}$ is a unit vector pointing in an arbitrary direction\cite{aublin}; (iv) several sources modeled according to equation \ref{eq:source}, in the directions $(l,b)$: $(0\degree,0\degree)$ [$\sigma=7.0\degree$, $A=1.00$], $(320\degree,90\degree)$ [$\sigma=1.5\degree$, $A=0.05$],  $(320\degree,-40\degree)$ [$\sigma=0.5\degree$, $A=0.01$],  $(220\degree,10\degree)$ [$\sigma=3.0\degree$, $A=0.05$],  $(100\degree,-70\degree)$ [$\sigma=2\degree$, $A=0.10$],  $(240\degree,50\degree)$ [$\sigma=20\degree$, $A=0.05$],  $(350\degree,-80\degree)$ [$\sigma=6.0\degree$, $A=0.005$],  $(100\degree,50\degree)$ [$\sigma=30\degree$, $A=0.50$],  $(140\degree,-40\degree)$ [$\sigma=4.0\degree$, $A=2.00$] and  $(60\degree,50\degree)$ [$\sigma=3.0\degree$, $A=0.02$]. This last background pattern was included in the simulation because unknown sources might be present during an analysis tagged on a given source, and their effects must be evaluated. From this combination of source (signal) and background patterns (noise), it was obtained the events map. 

\par The events map resulting from the sum of the background patterns and the simulated source was convolved with gaussian and MHWF (orders 1, 2 and 3) filters. The amplification of the SNR ($\lambda$) is calculated by the expression
\begin{equation}
	\lambda = \frac{w_f/\sigma_f}{w_0/\sigma_0},
\end{equation}
where $w_0$ is the value of the central pixel associated to the source in the non-filtered source map, $w_f$ is the corresponding value in the filtered source map, $\sigma_0$ is the root mean square (RMS) of the non-filtered background map and $\sigma_f$ is the RMS of the filtered background map.

\par Aiming to study the impact of the number of events from the source and its intensity upon the filter, different number of events ($N_{evt}$) were simulated in the direction of the source, ranging from 10 events up to 1000. The amplitude of the gaussian source ($A$) was also varied, from $10^{-4}$ to $1$.

\section{Results}

\par
\begin{figure}
	\centerline{\psfig{file=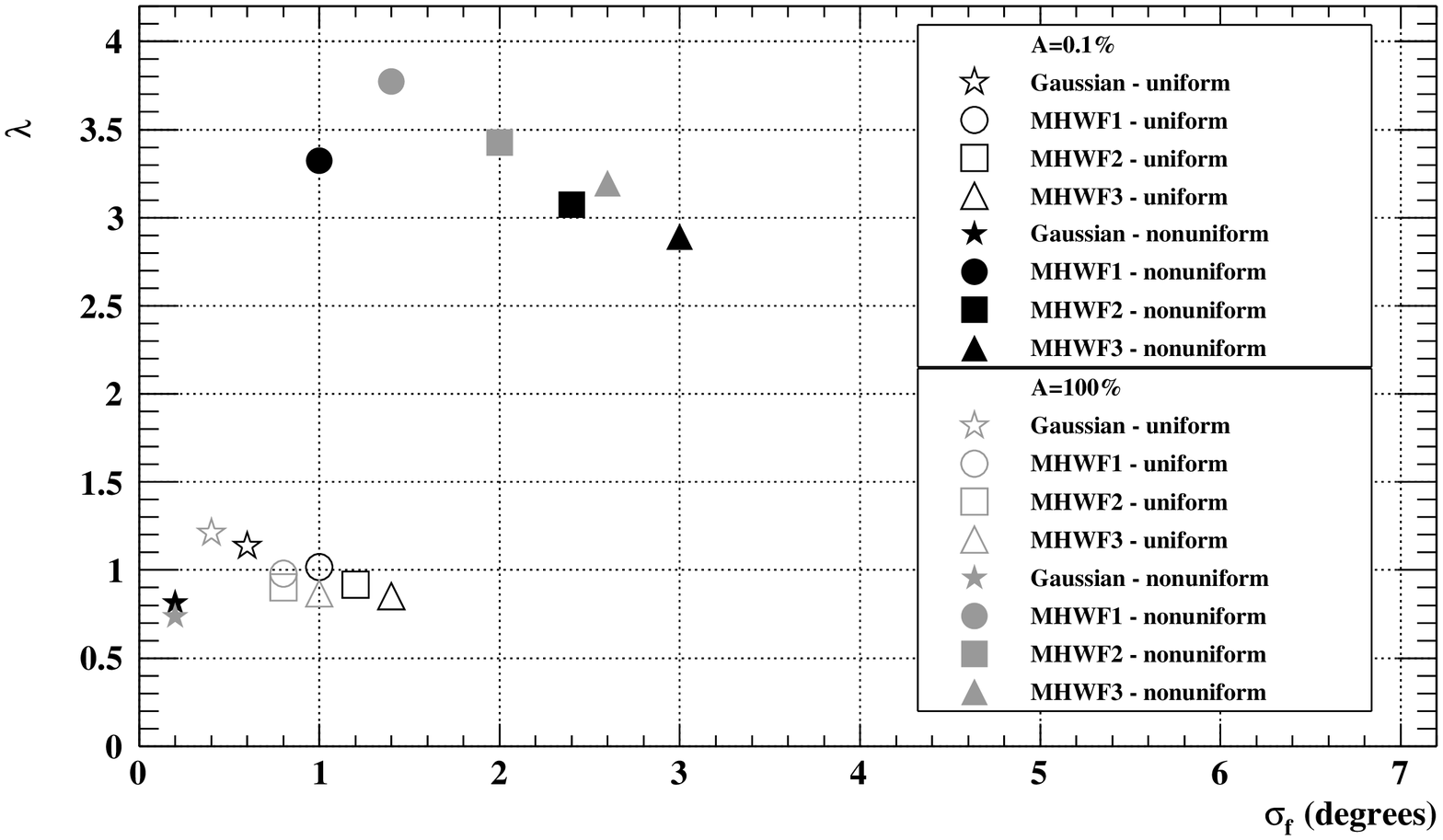,width=15cm}}
	\centerline{\psfig{file=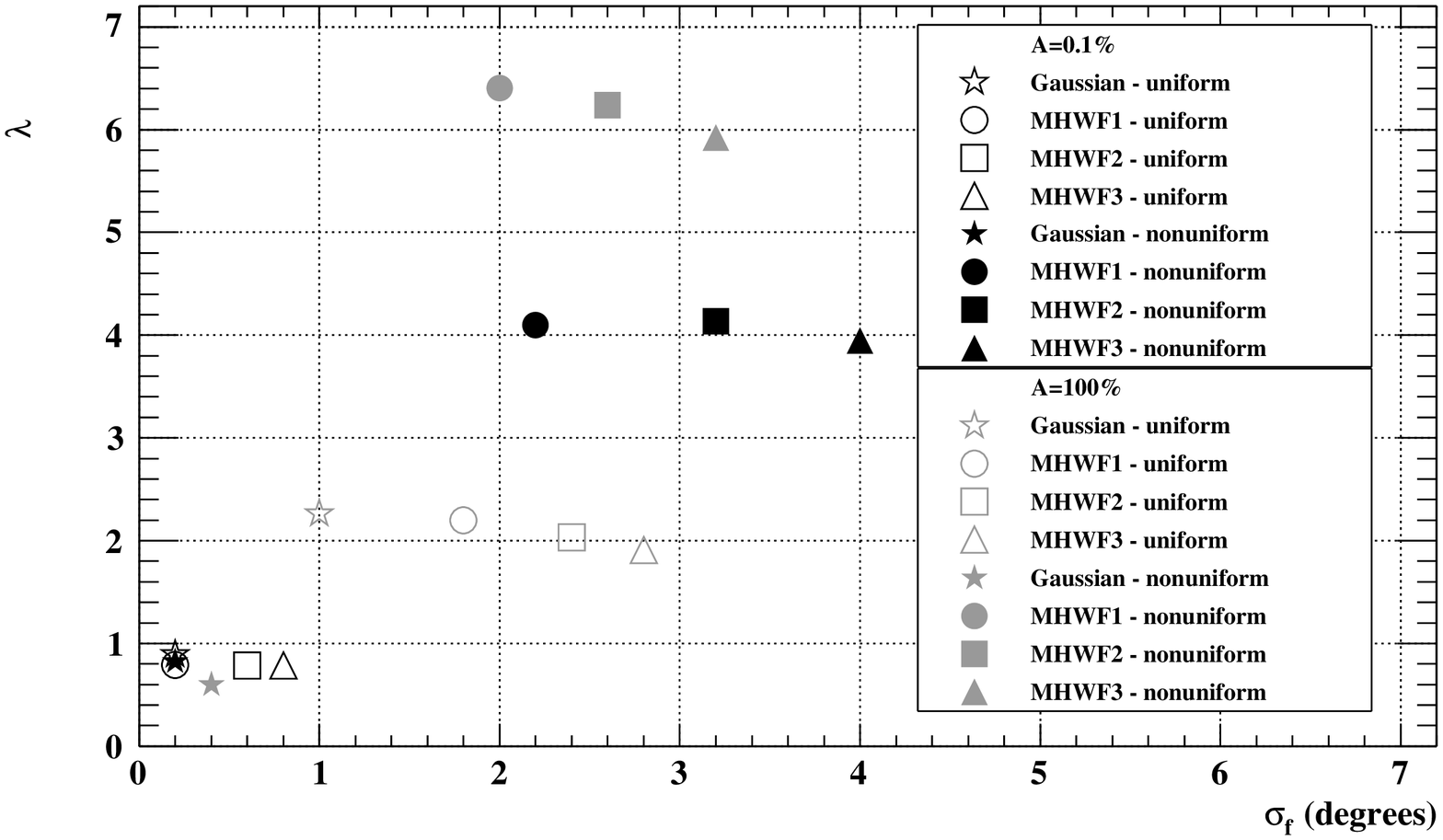,width=15cm}}
	\vspace*{8pt}
	\caption{Maximum amplification of the SNR ($\lambda$) as a function of the corresponding scale ($\sigma_f$). The graphs displayed refer to a fraction of events between the source and the background of $6 \times 10^{-6}$ (top) and $6.25\times10^{-4}$ (bottom).}
	\label{fig:evt}
\end{figure}

\begin{figure}
	\centerline{\psfig{file=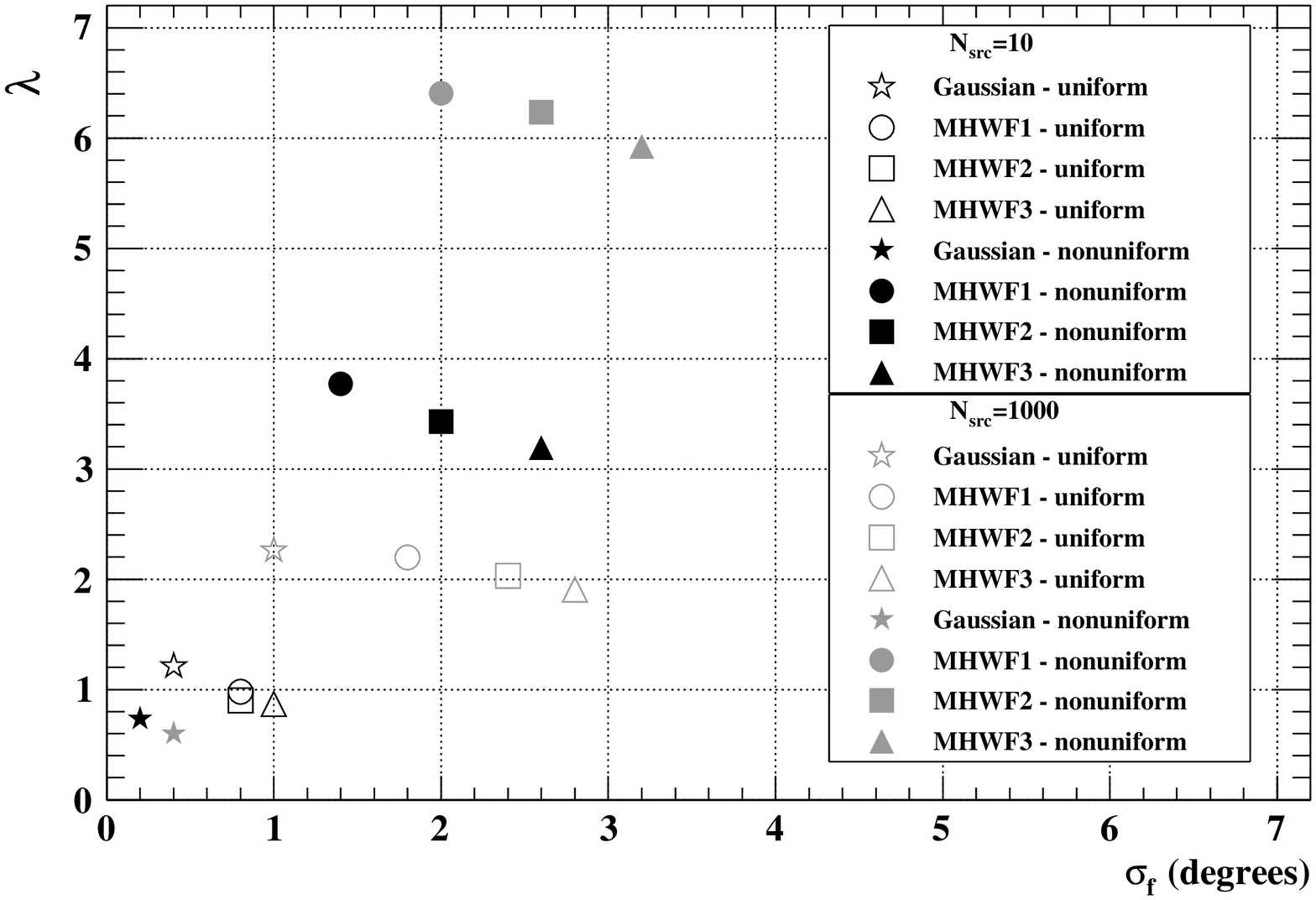,width=15cm}}
	\centerline{\psfig{file=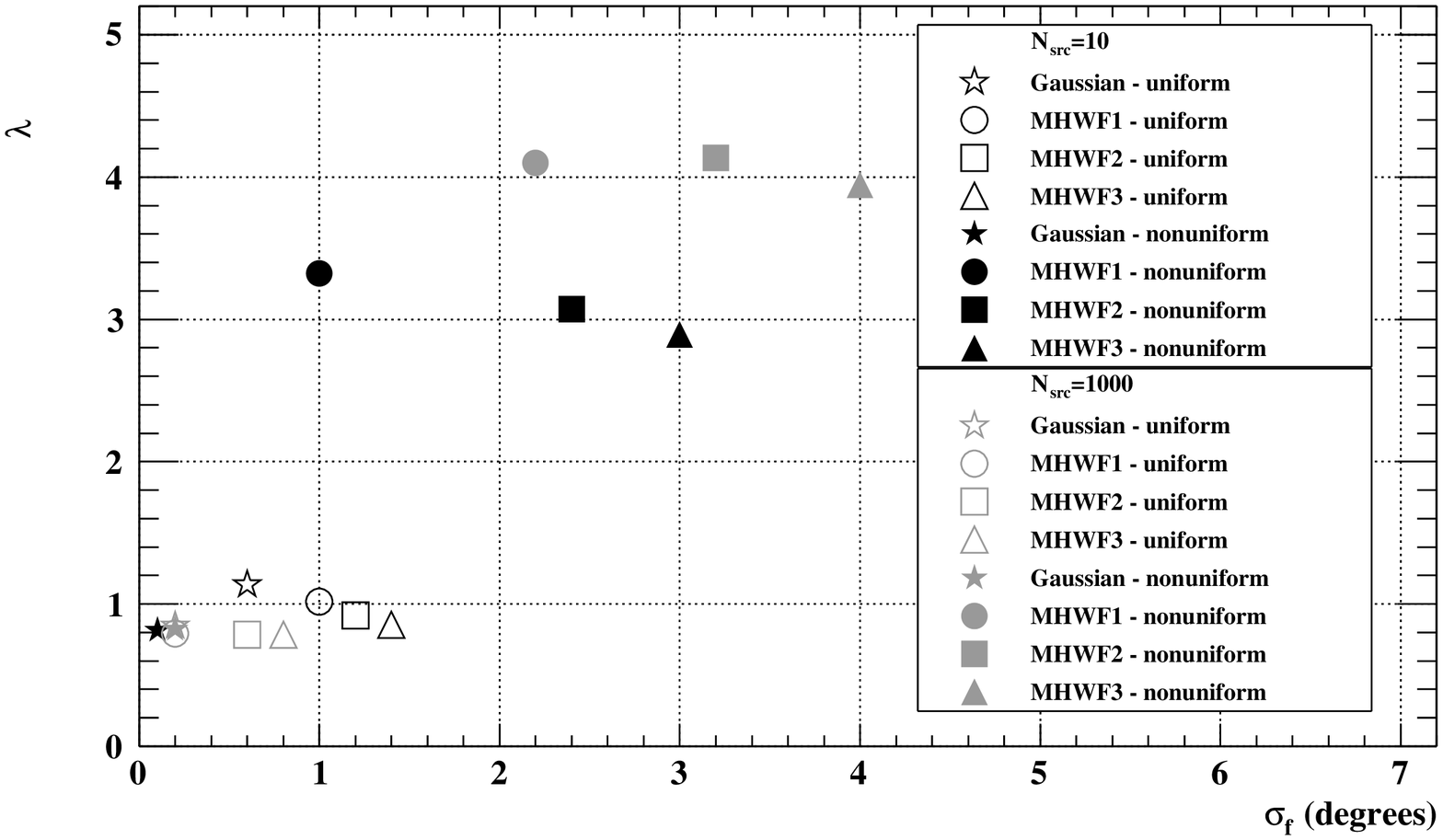,width=15cm}}
	\vspace*{8pt}
	\caption{Maximum amplification of the SNR ($\lambda$) as a function of the corresponding scale ($\sigma_f$). The graphs displayed refer to an amplitude A=$10^{-4}$ (top) and A=1 (bottom).}
	\label{fig:amp}
\end{figure}

 In figures \ref{fig:evt} and \ref{fig:amp} it is shown the maximum amplification of the SNR, as a function of the corresponding scale. In the case of figure \ref{fig:evt}, $N_{evt}$ is fixed and $A$ is varied. For figure \ref{fig:amp}, $A$ is fixed and $N_{evt}$ from the source is varied. 

\par From figure \ref{fig:evt} it can be noted that the uniform exposure acts like a constraint for $\lambda$, and that the gaussian has a slightly better performance than the MHWF filters. The maximum amplification for nonuniform exposure is clearly achieved by using MHWF filters, whereas the gaussian filter has amplification close to 1, which means no amplification. Comparing the two graphs in figure \ref{fig:evt}, in the bottom graph the amplification is greater, which seems reasonable since the number of events in this case is 100 times greater.

\par In figure \ref{fig:amp} it can be seen the behavior of the filters when $A$ is varied. For the uniform exposure  the amplification of both the gaussian and the MHWF filters are low, but the gaussian has a slightly better performance. Comparing the top and bottom graphs in figure \ref{fig:amp}, it is clear that the amplification achieved by the filters is proportional to $N_{evt}$. Also, when there is an acceptance, the gaussian filter does not provide a good amplification of the SNR, which can be achieved by using the MHWF.

\section{Conclusions}

\par In this work it was analyzed the performance of the gaussian and the MHWF filters  to detect point sources of cosmic rays embedded in a non uniform background, whose features are modulated both by the acceptance of the detector and the background patterns imposed to the incoming particles. Some parameters from the source such as the amplitude $A$ and the number of events $N_{evt}$ were varied, and the effects of theses changes on amplification of the SNR was studied.

\par The trivial conclusion is that the amplification achieved by these kernels is proportional to the source intensity parameters ($A$ and $N_{evt}$). It is interesting to notice that for a realistic case, i. e., a source with low amplitude and only a few events coming from its direction, the amplifications achieved are low for both filters.  The MHWF filters are more robust to these parameters and can provide a greater amplification of the SNR even if $N_{evt}$ is small.

\par Regarding the contribution of the acceptance of the experiment for the detection, it can be clearly seen that for uniform exposure the amplifications are smaller. In this case, the gaussian filter provides a slightly better amplification compared to the MHWF. However, for a more realistic case taking into account the nonuniform exposure of the detector, the MHWF filters always achieve a greater amplification. Also, they are more robust to low statistic of events, which makes them particularly useful for cosmic ray studies.

\end{document}